# On the effect of a two-rocks boundary on the propagation of nonlinear transients of temperature and pressure in deformable porous rocks.

E.Salusti[1], R. Droghei[2], R.Garra[3]

[1]INFN, Sezione Roma1, Dip. Fisica, Piazzale A. Moro, Roma, Italy

[2] ISAC-CNR, Via Fosso del Cavaliere 100, Roma,00100, Italy.

[3]Dipartimento di Scienze Statistiche, Università "La Sapienza", Piazza Aldo Moro, Roma 100.Italy

**Abstract**

We here analyze the propagation of transients of fluid-rock temperature and pressure through a thin boundary layer, where a steady trend is present, between two adjacent homogeneous rocks. We focus on the effect of convection on transients crossing such thin layer. In comparison with early models where this boundary was assumed a sharp mathematical plane separating the two rocks, here we show a realistic analysis of such boundary layer that implies a novel nonlinear model. Its solutions describe large amplitude, quick and sharp transients characterized by a novel drift and variations of the signal amplitude, leading to a nonlinear wave propagation. Possible applications are in volcanic, hydrologic, hydrothermal… systems as well as for deep oil drilling. In addition, this formalism could easily be generalized for the case of a signal arriving in a rock characterized by a steady trend of pressure and/or temperature. These effects, being proportional to the initial conditions, can also give velocity variations not particularly important. A further heuristic model has therefore been analyzed, i.e. assuming a pressure dependent rock permeability. In this way, a remarkable increase of the system velocities is obtained.

## 1. Introduction

Modeling of transients in fluid-saturated porous rocks, (Fig. 1) is of fundamental importance for a large number of applications in hydrological, volcanic, hydrothermal, hydrocarbon systems, deep oil drilling, fracking events... This is a classical problem: to analyze systems perturbed from a sudden arrival of pressure transients. Rice and Cleary (1976) envisaged the buildup of sources that trigger pressurized fronts through reactive porous horizons. McTigue (1986) considered also thermal processes in such theory. Bonafede (1991) focused on the role of fluid-rock energy equation. Natale et al. (1996) and Merlani et al. (2001) considered the nonlinear effect of the convection, which implies the presence of quick large amplitude fronts among the model solutions (Whitham, 1974).

These nonlinear models had many vivid applications as "La Fossa" crater in Vulcano (Aeolian Islands) in Italy (Natale, 1998), "The Geysers" in California (Moore and Gunderson, 1995; Natale et al., 1999), the Karymsky volcano in Kamchatka analyzed by Chirkov (1975), the submarine eruption off the Izu Peninsula in Japan (Notsu et al., 1991; Garcia et al., 2000). On a more theoretical ground Merlani et al. (2001) have considered also the parameters variability due to rock deformation/fracturing at the arrival of a large amplitude front, when the fluid approaches the fracture pressure region in a strain/forcing graph (Fig. 2). In these early analyses, the boundary between the "source" and an "adjacent" rock often has been assumed just as a mathematical plane characterized by rock parameter discontinuities.

The purpose of this paper is to examine in detail the transient evolution in such two-rock boundary layer of thickness $\psi$, seen as a thin region of continuous transition for fluid-rock temperature $T$ and pore pressure $P$, where a steady trend of $T$ and $P$ is present. We moreover stress that this approach can easily be generalized for the case of a source rock in presence of continuous steady $T$ and $P$ trends in the adjacent rock.

A novel nonlinear model has therefore been obtained, whose solutions determine a transient drift, and amplitude variations, during the transient propagation. We moreover follow the interest of Merlani et al (2001) for large amplitude transients, and discuss a novel model where we consider a pressure dependent permeability which gives more quick evolutions.

The paper is organized as follows. We first define early model equations about thermo-poro-elastic transients (Section 2) and their complexities and uncertainties in Section 3. The nonlinear model is investigated in Section 4. The solutions for the case of initial pressure trend is in Section 5. An analysis of the characteristic rock parameters is provided in Section 6. A model of the rock permeability changes for a strong pressure transient is discussed in Sections 7 and 8. A final discussion of our results is in Section 9.

**2. The thermo-poro-elastic equations for *P* and *T*.**

We quickly recall early models of a *P-T* front in *1-D* moving in a thin layer of thickness $\psi$ between two fluid saturated rocks, as a boundary layer of an aquifer that elastically reacts with an adjacent homogeneous rock (Fig. 1). In this *1-D* choice, the computations are easier, since the stress $\sigma_{ij}$ is constant (McTigue, 1986). This can hold for a two half-horizon schematization and similarly for a radial propagation from a small spherical source or for a cylindrical propagation from a perforated segment of a borehole, thus forming a segment source. In the half-horizon version, a boundary layer around $z \approx 0$ corresponds to the aquifer-rock boundary where temperature and pore pressure of the "*source*" rock are supplied by water circulating in a steady regime through the aquifer (Natale et al., 1999).

To describe such flows McTigue (1986) considered also a fundamental relation between the main quantities in such dynamics, namely *P* and *T*.

$$\left(\frac{\partial}{\partial t} - k^* \nabla^2\right)\left(\sigma_{kk} + \frac{3P}{B^*}\right) = \frac{3}{B^*}\left(\alpha^* \nabla^2 T + \alpha \frac{\partial T}{\partial t}\right), \tag{1}$$

where we call

$$k* = \frac{2K_f G}{\mu}\left[\frac{B*^2(1-v*)(1+v_u*)^2}{9(1-v_u*)(v_u*-v*)}\right] > 0,$$

$$\alpha* = \frac{4Gk*B*(1+v*)\alpha_m}{9(1-v*)} > 0,$$

$$\alpha = \frac{2GB*^2(1+v*)(1+v*_u)}{9(v*_u-v*)}\phi(\alpha_f - \alpha_m) > 0.$$

Here $B*$ is the Skempton parameter, $G$ is the shear modulus, $\sigma_{ij}$ is the stress tensor, $v*$ the drained Poisson ratio and $v_u*$ the undrained Poisson ratio, $\alpha_m$ ($\alpha_f$) the volumetric thermal expansion coefficient for the solid (fluid), $K_f$ is the medium permeability, $\phi$ the porosity and $\mu$ the fluid viscosity (see Table 1 and 2).

Equation (1) implies that for isothermal problems is the classical diffusion equation that rules the time evolution of $P$. For example, the evolution of pressure-induced micro-earthquakes are considered as an isothermal process, but with a pressure dependent permeability (Shapiro and Dinske, 2009). In turn, if temperature gradients are present, as often happens in real problems, the pressure evolution is ruled just by the gradients of $T$. The opposite also holds: in a constant pressure matrix, a quick thermal transient evolves as a mere diffusion.

In the general case, this equation is important since it interconnects strictly the evolutions of $T$ and $P$ for any initial input. In addition, since in a *1-D* geometry the stress $\sigma_{ij}$ is constant (McTigue, 1986) the equation (1) becomes (Table 1)

$$\frac{\partial P}{\partial t} - \alpha\frac{\partial T}{\partial t} - k*\frac{\partial^2 P}{\partial z^2} - \alpha*\frac{\partial^2 T}{\partial z^2} = 0. \tag{2}$$

As a second relation, following Bejan (1984) and Bonafede (1991) we assume a thermal balance

$$\frac{\partial T}{\partial t} = \frac{K_T}{\phi\rho_f c_f + (1-\phi)\rho_m c_m}\nabla^2 T - \frac{\rho_f c_f U \cdot \nabla T}{\phi\rho_f c_f + (1-\phi)\rho_m c_m} -$$

$$-\left(\frac{1}{\rho_f c_f + (1-\phi)\rho_m c_m} + \frac{\mu}{K_f}\Phi\right)U\cdot\nabla P =$$

$$= \frac{K_T}{\phi\rho_f c_f + (1-\phi)\rho_m c_m}\nabla^2 T + \frac{\rho_f c_f K_f}{\mu[\phi\rho_f c_f + (1-\phi)\rho_m c_m]}\nabla T\cdot\nabla P +$$

$$+\left(\frac{K_f}{\mu[\rho_f c_f + (1-\phi)\rho_m c_m]} - \Phi\right)(\nabla P)^2 = D\nabla^2 T + B\nabla P\cdot\nabla T + Y*(\nabla P)^2, \tag{3}$$

where we call

$$B = \frac{K_f \rho_f c_f}{\mu[\phi\rho_f c_f + (1-\phi)\rho_m c_m]} > 0,$$

$$D = \frac{K_T}{\phi \rho_f c_f + (1-\phi)\rho_m c_m} > 0,$$

$$Y^* = \frac{K_f}{\mu[\phi \rho_f c_f + (1-\phi)\rho_m c_m]} - \Phi,$$

Here $K_T$ is the average thermal conductivity, $\rho_m$ is the matrix density, $c_f$ and $c_m$ are the fluid and the rock heat capacity, respectively, and $U = -\frac{K_f}{\mu}\nabla P$ is the Darcy velocity (Tables 1 and 2).

The equation (3) is the classical heat conservation law but with two nonlinear terms, the convection $B\nabla P \cdot \nabla T$ and the mechanical work rate $Y^*(\nabla P)^2$ (Bejan, 1984; Bonafede and Mazzanti, 1997; Merlani et al., 2001). These terms can appear small quantities but they are related to auto-interacting effects, i.e. when a parcel of fluid flows towards a different point it also carries its $P$ and $T$ and eventual pollutants…. In turn, these are mathematically nonlinear unbounded functions, which can have explosive amplitudes also for small $B$ or $Y^*$. A balance of diffusion, convection and mechanical work rate therefore governs the evolution of the rock-fluid temperature.

## 3. The equations complexities and uncertainties.

Much care should be given to the physical meaning and experimental evaluation of the coefficients appearing in equations (2) and (3). In particular, the coefficient $Y^*$ should be treated with care: indeed the work made by $P$ increases the rock heat and therefore in (3) we have $Y^* > 0$ (Bejan, 1984). Nevertheless, if the perturbation gives rock deformations, fracturing or some kind of irreversible "*change of state*" in the rock, some energy, and heat, can be extracted from the matrix (Appendix A). In this case we can have that $Y^* \to Y^* - \Phi = Y < 0,$ for a suitable $\Phi > 0$ that takes into account these dissipative effects (Gross and Seelig, 2006). Classical cases are the energy dissipated in the rock to create new fractures (Philipp et al., 2013), frictional heating during an earthquake (Rice, 2006) or also the energy dissipated by the viscous fluid moving inside the fractures (Detournay and Garagash, 2003).

All this explains why a realistic determination of the rock parameters $B, D, Y, \Phi$... for a deep rock is not a simple challenge. Moreover these rock parameters moreover are quantities poorly known, to be checked by comparing with other information. Thus, the above equations, depending from many empirical coefficients often rather poorly known, must be considered critically.

We here mention a further point: the rock parameters often considered as constants, in reality can be perturbed during the transient impact. Thus one has to take into account that significant *P-T* jumps in the matrix can build variations of the matrix parameters (Appendix A; Fisher et al., 2002; Philipp et al., 2013).

In more detail, in a hydrocarbon or geothermal system the arrival of a strong *P-T* transient can produce rock deformations and/or hydrofractures, which can propagate until forming interconnected fracture systems. In such situation variations of Poisson's ratio, Skempton parameter, shear modulus, rock thermal expansivity, rock density and rock heat capacity are due to variations of the porosity and thermal coefficients, i.e. rather bounded quantities (Bonafede ,1991). Variations of these quantities can consequently be less important over, say, ranges of temperature and pressure around $T \approx 100\ °C$ and $P \approx 10^7\ Pa$, i.e. characteristic values for volcanic processes (Zencher et al., 2006; Fisher et al., 2002). On the other hand, the permeability $K_f$ can be severely affected by the arrival of a strong $P$ transient (Gross and Seelig, 2006; Shapiro and Dinske, 2009). Indeed, such arrival can lengthen fractures and cracks, interconnect many minor cracks until a thoroughgoing net is created (Appendix A). Thus in Section 6 some parameters as $K_f$ (Table 1) and the a corresponding $k^*, \alpha^*, B, Y, \Delta, \Sigma, V...$ in Table 2, are considered to vary at the arrival of a strong front.

We also remark that these problems can be characterized by widely different parameters. For example, intrusive dykes can have lengths as 1-10 $Km$ and widths of 1-10 $m$ (Zencher et al., 2006). In contrast, injections of viscous fluid in a matrix can give fractures with a length of 1-10 $m$ with a width of 1- 10 $cm$ (Detournay and Garagash, 2003). Another interesting case is the thermal diffusion in an earthquake slip where a very small thickness between 1.0 $\mu m$ and 1.0 $mm$ has been considered (Rice, 2006; Rempel and Rice, 2006). The corresponding values of the boundary thickness $\psi$ must be much smaller and we here tentatively assume that $\psi$ is about 10% of the above widths.

## 4. The constant-permeability models.

In synthesis, from (2) and (3) we have a system of two equations in *1-D*

$$\frac{\partial P}{\partial t} = \alpha \frac{\partial T}{\partial t} + k^* \frac{\partial^2 P}{\partial z^2} + \alpha^* \frac{\partial^2 T}{\partial z^2} \ , \tag{4}$$

$$\frac{\partial T}{\partial t} = B \frac{\partial T}{\partial z} \frac{\partial P}{\partial z} + Y \left(\frac{\partial P}{\partial z}\right)^2 + D \frac{\partial^2 T}{\partial z^2} \ . \tag{5}$$

Rice and Cleary (1976), McTigue (1986), Bonafede (1991), Natale and Salusti (1996) among many others, consider the boundary/initial conditions of a "*source*" matrix ($T = T_0 + T_I$ and $P = P_0 + P_I$ for $z < 0$ at $t \approx 0$) and an adjacent matrix ($T = T_0$, $P = P_0$ for $z > 0$ at $t \approx 0$) with $P_0$, $P_I$, $T_0$, $T_I$ constants. It is clear that such conditions hold as long as the upstream matrix, the "*source*", is so wide that the transient does not affect its $P$ and $T$ evolution.

In order to have a first idea of the values here analyzed, we remark how in the literature often are examined problems where $P_I \approx 10^7\ Pa$ and $T_I \approx 100\text{-}1000\ °C$ for earthquakes, volcanic or geothermal systems (Bonafede ,1991); $P_I \approx 10^7\ Pa$ and $T_I \approx 60\ °C$ for induced micro seismicity analyses (Fisher et. al., 2002; Shapiro and Dinske, 2009); $P_I \approx 10^4\ Pa$ and $T_I \approx 10\ °C$ in the McTigue (1986) study about nuclear waste disposals and $P_I \approx 50\ MPa$ and $T_I \approx 200\ °C$ for earthquake slip analysis (Chester et al., 2005; Rice, 2006).

We can now define in full detail the problem that we are analyzing: since a sharp jump, as initial condition for such transients is not a realistic assumption, we analyze the *P* and *T* evolution considering a thin layer between the source and adjacent rocks, and call $\psi$ its thickness. In this layer, a steady trend of *P* and *T* has a dynamical effect on a transient evolution, which requires a more complex novel model. In addition, such formalism for a large $\psi$ can simulate a steady trend of *P* and/or *T* in the adjacent rock.

In these early analyses, the model solutions are functions of $z^2/t$ characteristic of pure diffusion problems. Following such remark Merlani et al. (2001) studied the "*symmetry*" properties of (4) and (5) and obtained as explicit solutions the "*rigid wave translations*", i.e. $G(z - v t + c)$, or the "self-similar" solutions $F(z^2/t + c)$, where *v* and *c* are constants. The first case looks rather artificial since only rigid moving profiles can be obtained, thus we investigate a self-similar case. We therefore assume a particularly simple *ansatz* (Merlani et al., 2011)

$$k*\frac{\partial^2 P}{\partial z^2} + \alpha*\frac{\partial^2 T}{\partial z^2} = \frac{d f(t)}{d t} \quad . \tag{6}$$

From (4) the previous equation is equivalent to

$$P(z,t) = \alpha T(z,t) + g(z) + f(t) \quad , \tag{7}$$

a position related to the initial/boundary conditions, which must be checked in the final solutions. Garra et al. (2015) analyzed a somehow similar problem by applying a "fractional" memory formalism.

The novel Darcy velocity from (7) is

$$U(z,t) = -\frac{K_f}{\mu}\frac{\partial P}{\partial z} = -\frac{K_f}{\mu}\frac{\partial}{\partial z}\left[\alpha T(z,t) + g(z)\right] \quad . \tag{8}$$

In our study $g(z) = P(z, 0) - \alpha T(z, 0)$ plays a key role in such transient evolution for $0 < z < \psi$. As an example, for *z* upwards the Earth thermal gradient is about $-3 \cdot 10^{-2}$ °C /m and that of hydrostatic pressure is about $-10^4$ Pa/m (Table 2).

To solve (4) and (5) we define $\Delta = Y \alpha^2 + \alpha B$ and $\Sigma = 2 Y \alpha + B \approx 2\Delta/\alpha$ (Table 2). From (5) and (7) we obtain the following Burgers-like equation

$$\frac{\partial T}{\partial t} = D\frac{\partial^2 T}{\partial z^2} + \Delta\left(\frac{\partial T}{\partial z}\right)^2 + \Sigma\frac{d g}{d z}\frac{\partial T}{\partial z} + Y\left(\frac{d g}{d z}\right)^2, \tag{9}$$

(more exactly its *z*-derivative is a Burgers equation, with a drift and a forcing). About equation (9) Whitham (1974) defines a Reynolds number

$$R = \frac{(Y\alpha^2 + \alpha B)\cdot T_1}{D} = \frac{\Delta \cdot T_1}{D}, \tag{10}$$

that characterizes its solution to be shock waves or classical diffusion solutions (Fig.3). It is important to stress how **R** is proportional to the initial value $T_I$, different from a pure diffusion.

## 5. The solutions for small and large amplitude transients.

From the structure of equations (4) and (5) or (9) we see that if $T(z)$ is a solution, then also $T(-z)$ and $T(z) + const$ are solutions with different boundary conditions. In addition, a change of sign of $\Delta$ implies a change of sign for the novel solution, if $Y$ and $\Sigma$ are nil. The role of $T_I$, $P_I$ and $\Delta$ are therefore essential since different physical settings are related to their signs. We moreover note how $\Delta$ is positive for mild transients, since eventual fracturing processes have little importance; while for very large $P_I$ or $T_I$ an eventual "*change of state*" is possible, then allowing a negative $\Delta$. We will therefore discuss both cases, of positive or negative $\Delta$.

We are well aware that the solutions of equation (9) can reach a high level of complexity for irregular behaviors of $P(z,0)$ and $T(z,0)$. In order to give an intuitive example of the solutions of the thin boundary model, we here discuss a steady trend with a negative linear gradient of $P(z,0)$, but with $T(z,0)$ constant. In the $0 - \psi$ interval we therefore have

$$g(z) = P(z,0) - \alpha T(z,0) = P_0 - \alpha T_0 - \Gamma z/\psi, \tag{11}$$

and thus $\dfrac{d}{dz} g = -\Gamma/\psi$ for $\Gamma > 0$. This $\Gamma$ is the steady pressure difference for $\psi > z > 0$. As initial temperature we consider only a sudden transient jump $T_I$ at $t \approx 0$, as those above described.

A quadratic version of (11) demonstrates that the larger effects are in zones with larger $dg/dz$ (not shown in this paper). The effect of a more irregular $g(z)$ is also rapidly sketched in Appendix B.

***The positive $\Delta$ case.*** We now analyze the effect of an initial continuous steady gradient of pressure at the arrival of just a temperature jump $T_I$. We assume rather small initial inputs, such that $Y, \Delta, \Sigma, \Gamma$ are positive (Table 2). In this simple case equation (9) for $0 < z < \psi$ becomes

$$\frac{\partial T}{\partial t} = D \frac{\partial^2 T}{\partial z^2} + \Delta \left(\frac{\partial T}{\partial z}\right)^2 - \Sigma \frac{\partial T}{\partial z} \frac{\Gamma}{\psi} + Y \left(\frac{\Gamma}{\psi}\right)^2 \qquad t > \varepsilon, \tag{12}$$

where $\varepsilon$ is a small time delay to take into account all the complex processes related to a quick arrival of a real transient (Appendix A). If again $\mathbf{R} = |T_I \Delta|/D > 8 - 10$, in (12) one can disregard $D$ (Whitham, 1974).

Moreover the equation (9) has constraints (Appendix C) that define a transient front at $z = z_B(t)$. Indeed, the solution for $t > \varepsilon$ is

$$T(z,t) = T_0 + T_I \qquad\qquad z < 0,$$

$$T(z,t) = T_0 + T_I - \frac{(z+Vt)^2}{4\Delta t} + Y\left(\frac{\Gamma}{\psi}\right)^2 t \qquad 0 < z < z_B(t) \approx -Vt + \sqrt{4\,T_I \Delta\,t} \quad , \qquad (13)$$

$$T(z,t) = T_0 \qquad z > z_B(t),$$

with a drift velocity $V = (\Sigma\Gamma)/\psi = (2\Gamma\Delta)/(\alpha\psi) > 0$ which decreases the front velocity (Fig 4). From Fig. 4 we see how temperature quickly increases until about $T_0 + T_I$, not at the arrival of the front $z_B(t)$ but after a time delay, then the temperature remains rather stable, about $T_0 + T_I$.

Equation (13) also shows that for a fixed time the temperature decreases for increasing $z$.

The velocities $V$ are positive but rather small, around $10^{-8}$ in SI. Since $Y \approx 10^{-21}$ also $Y\Gamma^2/\psi^2$ is small, about $10^{-9}$ for sandstones and $10^{-12}$ for Tennessee marble.

On the other hand, for very small $\psi \approx 50\ \mu m$ (Chester et al., 2005), this velocity $V = \Sigma P_I / \psi$ can reach higher values, also $\approx 10^{-4}$ in SI. In such a context Rice (2006) states that "*the earthquake data set for fracture energies can be fit to predictions of a model involving slip on a much thinner zone, even slip on a mathematical plane. It is, nevertheless, presently uncertain whether broad zones of ultracataclastic gouge, up to several tens of millimeters width, participate in seismic shear, or whether extreme localization is the rule even if such localized zones may have in some cases evaded detection.*"

To clarify the meaning of **R**, the diffusive velocity of (9) is $\approx \sqrt{D/t}$ while the nonlinear velocity is $\approx \sqrt{|T_I \Delta|/t}$. The physical characterization of ratio between the two velocities therefore is $\sqrt{|T_I \Delta|/D} \approx R^{1/2}$.

From (13) we moreover have that at the second interface $z = \psi$ the temperature evolves as

$$T(\psi,t) = T_0 + T_I - \frac{(\psi + Vt)^2}{4\Delta t} + Y\left(\frac{\Gamma}{\psi}\right)^2 t \quad , \tag{14}$$

and for large times $T(\psi,t) \to \left[Y\left(\frac{\Gamma}{\psi}\right)^2 t - \frac{V^2}{4\Delta}t\right] \approx -10^{-6} t$, i.e. its decrease is proportional to $t$.

Such $T(\psi,t)$ can play the role of new initial/boundary conditions for the transient evolution in the other "*adjacent*" matrix. In addition to it, the time $t^*$, necessary to reach the second rock, for small $Y$ can be computed from

$$\psi = -Vt^* + \sqrt{4\Delta T_I/t^*}\ , \tag{15}$$

and for small $V$ and $Y$ we obtain that $t^*$ is proportional to $\psi^2/T_I$, very small for a small $\psi$.

From (7) the corresponding pressure for $t > \varepsilon$ is given by

$$P(z,t) = P_0 + P_I + \alpha T_1 \qquad z < 0$$

$$P(z,t) = P_0 - \Gamma \frac{z}{\psi} + \alpha \left( T_I - \frac{(z+V\,t)^2}{4\Delta\,t} + Y \left(\frac{\Gamma}{\psi}\right)^2 t \right) \quad 0 < z < z_B(t) \approx -Vt + \sqrt{4\,(T_I - \frac{\Gamma z_B}{\alpha \psi})\Delta\,t}$$
(16)

$$P(z,t) = P_0 - \Gamma z/\psi \qquad\qquad z_B(t) < z < \psi$$

We stress the relevance of these solutions where $T$ and $P$ satisfy exactly the assumptions (6) - (7), our *ansatz*.

***The negative $\Delta$ case***. For larger external signals one can also have a similar problem but with negative $Y, V, \Sigma$. This characterizes the transient evolution when the transient $P$ and $T$ are much larger, near the rock fracture region in Fig 2. The solutions are a sharp signal with the same front $z_B(t)$ since for $t > \varepsilon$ we have (Fig 5)

$$T(z,t) = T_0 + T_I \qquad\qquad z < 0$$

$$T(z,t) = T_0 - \frac{(z+V\,t)^2}{4\Delta\,t} + Y\left(\frac{\Gamma}{\psi}\right)^2 t \qquad 0 < z < z_B(t) \approx -Vt + \sqrt{4/\,T_I\Delta/t} \quad (16a)$$

$$T(z,t) = T_0 \qquad\qquad z > z_B(t)$$

but in this case the front is characterized as $T(z_B(t), t) = T_0 + T_I$.

It has to be remarked that for this negative $\Delta$ one has a negative $V$, which in turn increases the front velocity. In addition, the transient temperature at the front arrival is a sharp jump as large as $T_0+T_I$ and then decreases, till reaching $T = T_0$: this solution (16a) is just a quick transient. Again, these solutions of $T$ and $P$ satisfy exactly the assumption (6) - (7).

**6. The parameters values for some rock examples.**

The parameters $V, \Delta, \Sigma$..... in Table 2 have a fundamental role on such transient evolutions. For Charcoal granite, Tennessee marble or Westerly granite, the permeability $K_f$ is of the order of magnitude of $10^{-19}$ in SI. The diffusion parameter is often $D \approx 10^{-6}$ while $\alpha \approx 10^5$ in SI but with remarkable variations. For these rocks $\mathbf{R}/T_I$ is therefore smaller than $10^{-3}$ and the transients are essentially due to the diffusion. We can also see that the velocity $V \approx 10^{-10}$ in SI: therefore in these rocks the convection cannot play an important role since the diffusive effects mainly drive the system evolution (Table 2).

Sandstones in turn, and Berea sandstone in particular, have larger values for $K_f$ and thus $\mathbf{R}$. For these rocks the main nonlinear dynamical effects (namely $\mathbf{R} > 8-10$) can be seen if $T_I$ is larger than, say, 10 -100 $^0C$ and thus the $P$ and $T$ shock waves are easier to be found in these rocks. We moreover stress how the nonlinear front velocity $\approx 10^{-4}/\,t^{1/2}$ is much larger than $V \approx 10^{-7}$ in SI (Table 2) but it also has a rather quick decrease. Indeed, the ratio of such two velocities is about $10^3/(t^{1/2})$, assuming $\Gamma/2\psi = 10^6$ and $T_I = 100°C$.

**7. The equations of the large pressure model, with pressure-induced permeability.**

In the preceding sections we analyzed the dynamics in the boundary layer between two rocks at the arrival of a nonlinear transient, but the velocities for a small $\psi$ are not particularly impressive. Thus we take into account that significant *P-T* jumps in the matrix (Fig 2) can generate variations of the matrix parameters, in particular permeability (Bonafede, 1991; Shapiro and Dinske, 2009; Hummel, 2013). To this purpose we analyze the case of a $K_f(P)$ for large amplitude transient with *P* and *T* near the fracture value in Fig 2 (Appendix A; Fisher et al., 2002). We thereafter discuss a novel heuristic model with $K_f = K_f(P)$. We call $\pi(P) = K_f(P)/K_f$ the fluid pressure-dependent permeability over the constant low-pressure permeability. We moreover assume that for large amplitude transient *V, Y, Λ, Σ* are negative quantities, as discussed in Section 5.

About early estimates of the pressure dependent permeability, Gangi (1978) discussed a model with $K_f(P) = k_0 [1+ (P/P_e) m]^3$ where $k_0$ is the "*laboratory*" permeability, $P_e$ is an effective modulus and *m* is a constant, with $0 < m < 1$. More recently, among other exponential approaches (Yilmaz et al., 1994), a polynomial $K_f(P)$ was estimated from the analysis of micro-earthquakes clouds in experiments of deep pressure explosions in Barnett Shale (Fisher, 2002). In particular Hummel (2013) found $\pi(P) = S_0 P^n$ with $5 < n < 7$, where for $n = 5$ was $S_0 = 5 \times 10^{-32}$, for $n = 6$ was $S_0 = 5 \times 10^{-39}$ and for $n = 7$ was $S_0 = 5 \times 10^{-45}$ in SI. The largest $\pi(P)$ increase thus is about $10^3$. A similar behavior was also found for micro-seismic data originating from the Horn River Basin (Hummel, personal communication).

The novel model equations for $K_f = K_f(P)$ therefore are

$$\begin{cases} \dfrac{\partial P}{\partial t} = \alpha \dfrac{\partial T}{\partial t} + k * \pi(P) \dfrac{\partial^2 P}{\partial z^2} + \alpha * \pi(P) \dfrac{\partial^2 T}{\partial z^2} \\ \dfrac{\partial T}{\partial t} = B \pi(P) \dfrac{\partial T}{\partial z} \dfrac{\partial P}{\partial z} + Y \pi(P) \left(\dfrac{\partial P}{\partial z}\right)^2 + D \dfrac{\partial^2 T}{\partial z^2} \end{cases} \quad (17)$$

where only $\dfrac{\partial T}{\partial t}$ and $\dfrac{\partial P}{\partial t}$ are multiplied by $\pi(P)$. For a general case, the solutions of this problem could be found numerically, nevertheless simple analytic estimates can have some practical interest.

**8. The solutions of the large pressure model with $K_f(P)$.**

In the case of strong initial conditions to have an estimate from (17) we approximate

$\pi(P) = 1 + S_0 (P-P_0)^n \approx S_0 (\alpha z^2/4\Delta[t+\varepsilon])^n = S_0 (\alpha/4\Delta)^n z^{2n} /[t+\varepsilon]^n =$

$$= S_0 \, (\alpha/8\Delta)^n \, \psi^{2n}/[t+\varepsilon]^n = D^*/[t+\varepsilon]^n \qquad (18)$$

where we consider the time $\varepsilon$ but disregard the $V$ and $P_I/\psi$ terms. This avoids mathematical pathologies and takes into account the delay due to all the complex processes related to a quick arrival of a real transient (Appendix A). As suggested from the above experiments of micro-earthquake clouds, we estimate from (18) that at the pressure maximum one has

$$10^3 \approx D^* \, \varepsilon^{-n} \qquad (19)$$

and thus $\pi(P) \approx 10^3 \cdot \left(\dfrac{\varepsilon}{\varepsilon+t}\right)^n$. We also approximate $z^2 \approx \psi^2/2$ namely the average of $z^2$ in the $0 - \psi$ layer, since $z^2$ for a thin boundary has variations less important than $1/t$. With such approximations $\pi(P)$ becomes just a function of time.

In synthesis the above field data allows to determine in (17), (18) and (19) an heuristic model of the effects of $K_f(P)$ on large amplitude transients in small space scales, which at least holds for experiments in this range of parameters. Thus we finally obtain

$$\frac{\partial P}{\partial \theta} - \alpha \frac{\partial T}{\partial \theta} - k^* \frac{\partial^2 P}{\partial z^2} - \alpha^* \frac{\partial^2 T}{\partial z^2} = 0, \qquad (20)$$

$$\frac{\partial T}{\partial \theta} - \left[ \Delta \left(\frac{\partial T}{\partial z}\right)^2 + \Sigma \frac{P_I}{\psi} \frac{\partial T}{\partial z} + Y \left(\frac{P_I}{\psi}\right)^2 \right] = 0, \qquad (21)$$

as the above discussed models but with $t \rightarrow \theta(t)$. From (20) and (21) this new time $\theta$, determined from (17), is

$$\frac{d\theta}{dt} = \pi(P) = 1 + D^*(t+\varepsilon)^{-n}. \qquad (22)$$

By imposing that $\theta(0) = 0$ we moreover obtain

$$\theta(t) = t + \frac{D^*}{(1-n)}(t+\varepsilon)^{(1-n)} - \frac{D^*}{(1-n)}\varepsilon^{(1-n)} > t. \qquad (23)$$

For large times $\theta(t) \rightarrow t + D^* \varepsilon^{-n} \approx t$ while for very small times a perturbative expansion gives $\theta(t) = t \, [1 + D^* \dfrac{(n+1)}{(n-1)} \varepsilon^{-(n+1)}] \approx t \, 10^3/\varepsilon$ (Fig 6).

The corresponding temperature is

$$T(z,t) = T_0 - \frac{(z+V\theta)^2}{4\Delta\,\theta} + Y\left(\frac{\Gamma}{\psi}\right)^2 \theta \qquad 0 < z < z_B(t) \approx -V\theta + \sqrt{4/T_1\Delta/\theta} \ . \qquad (24)$$

In synthesis, one has for large times the same evolution already analyzed, while for very short times a novel, enhanced time is $\theta(t) \approx t\,10^3/\varepsilon$, *as* it would be intuitively expected. In turn the new front velocity for $t > \varepsilon$ is

$$\frac{d\,z_B}{d\,t} \approx -V\frac{d\theta}{dt} + \sqrt{\frac{|\Delta|/T_I}{\theta(t)}}\ \frac{d\theta}{dt} \approx 10^3\left(-V + \sqrt{\frac{|\Delta|T_I}{\theta(t)}}\right)\left(\frac{\varepsilon}{\varepsilon+t}\right)^n . \qquad (25)$$

In comparison with the constant-permeability estimates for sandstones, in particular the Berea sandstone in Section 6, the novel solution (25) has both **R** and the drift $V$ about $10^3$ times larger. For $\Gamma/2\psi \approx 10^6$ and $T_I \approx 100\,°C$ the new drift velocity $V$ is $\approx 10^{-4}$. In turn the nonlinear velocity is only $\approx$ 30 times larger and approximately it is $10^{-3}\,(t)^{1/2}$ in SI. This consequently shows how for a variable permeability the drift velocity plays a more important role. Similar relations hold also for $T(\psi,t)$ in (14) and for $t^*$ in (15). In Appendix D the concomitant effect of a filter cake formation is quickly sketched.

**9. Discussion.**

A realistic analysis of the effect of a boundary layer between two fluid saturated rocks, on the propagation of transients of $P$ and $T$, is an important but rather complex problem. It is evident that it can have different physical meanings, which dynamics may evolve in many different ways. One indeed can consider a pure pressure flux or pure heat flux or a mixing of these two forcings; the source can have higher or lower temperature and pressure, temperature only, pressure only, with or without rock deformations, a steady trend of $P$ and $T$ is present...etc. Also nonlinear effects as convection or advection can eventually be considered. In addition, such boundary layer can be thin, and schematize the real boundary layer between two homogeneous rocks, or very large to simulate an adjacent rock with some $P$ and/or $T$ trends, two very different but formally similar problems.

We here consider in detail the case in which a homogeneous "*source*" matrix is more pressurized than another adjacent matrix, thus focusing on a $P$ and $T$ dynamics in a thin boundary layer of thickness $\psi$ between these two rocks. Such barrier is here treated as a region with continuous trends of pressure between the two matrices: its thickness $\psi$ can be 10 *m* or 1 *mm* depending by the particular physical problem considered. The effect of a sudden jump of temperature in the source rock on the evolution of a continuous pressure field in such boundary layer is here analysed: we focus on its nonlinear pressure propagation.

In more detail, we analyse a two equations model describing these cases where we also consider the nonlinear role of convection. Indeed we focus on the characteristics of the model solutions in relation to some crucial parameters as **R** (a Reynolds number ruling the effect of convection), $\Delta$ (that characterizes the velocity of a nonlinear front), $\alpha$ (relating strictly $P$ and $T$), $V$ (a drift velocity due to eventual steady trends). In particular, we show that among the model solutions, there are also particularly quick and sharp transients with a strict linear relation between $P$ and $T$, as discussed by Merlani et al. (2011).

Solutions of these two equations can be very complex for realistic $P$ and/or $T$ trends (Appendix B), thus we here analyse a case characterized by simple analytical solution, i.e. the arrival of a sharp temperature jump in presence of a steady linear pressure trend. The main result is that all this gives a rather small drift velocity between the two rocks and the solution is

$$T(z,t) = T_0 + T_I - \frac{(z+V\,t)^2}{4\Delta\, t} + Y\left(\frac{\Gamma}{\psi}\right)^2 t \qquad 0 < z < z_B(t) \approx -Vt + \sqrt{4\,T_1\,\Delta\,t} \qquad (26)$$

for $t > \varepsilon$.

We study also the effects of convection for the case in which a stronger impact gives some rock perturbation or deformation. This corresponds to a transient with $\Delta < 0$ and again a front $z_B(t)$ as

$$T(z,t) = T_0 - \frac{(z+V\,t)^2}{4\Delta\, t} + Y\left(\frac{\Gamma}{\psi}\right)^2 t \qquad 0 < z < z_B(t) \approx -Vt + \sqrt{|4\,T_1\,\Delta/t|} \quad . \qquad (27)$$

From these analyses, we however found that for volcanic, hydrothermal... problems the drift velocity $V$ is rather small**,** only for an earthquake slip it can be more important (Rice, 2006). We therefore investigate a case with an also larger external impact, such that the transient pressure increases the rock permeability $K_f$. This is the case of a $K_f(P)$ reaching values $\approx 10^3\, K_f$ (Shapiro et al., 2006). The novel solutions have only a different time $\theta(t)$, which gives a large increase of the system velocities, in particular for the drift velocity. For sandstones, this front velocity is roughly about $10^{-2}/\sqrt{t}$ in SI while the positive drift velocity $V$ is $\approx 10^{-4}$ in SI, which shows the complex dynamics of these problems.

**Acknowledgments**

We must thank Prof. Abbasbandy and Dr. Hashemi for help in an early stage of this study. This research is partially supported by the Italian RITMARE Flagship Project.

**Appendix A. Summary remarks about hydro-fractures in porous rocks.**

In many articles dykes, mineral veins, joints and man-induced hydraulic cracks generated by fluid overpressures are called hydro-fractures (Philipp et al., 2013). Such events can influence the rock permeability in reservoirs of oil, gas, geothermal water and groundwater. Indeed various analytical and numerical models show that such hydro-fractures pressure jumps often give very high crack tip tensile stresses (Shapiro et al., 2006). Then these hydro-fractures propagate until forming an interconnected net of fractures. In turn, these contribute to enlarge the rock permeability.

Field observations show how in heterogeneous or vertically layered rocks hydro-fractures can be arrested at a horizontal layer boundary, thus vertically interconnected networks are not frequent (Philipp et al., 2013). This often happens to hydro-fractures with discontinuities (including contacts) or stiffness changes between horizontal layers and stress barriers, i.e. the local stress there contrasts any kind of vertical hydro-fracture propagation. Thus, hydro-fractures are mostly confined in stratabound layers where often a net of fractures forms. Such phenomena can in turn largely contribute to enlarge the fluid reservoirs permeability.

**Appendix B. On the effect of strongly varying initial conditions at the two-rock boundary.**

A very different problem is considered here, we analyze the assumption for the initial pressure in presence of trend

$$P_0(z) = P_0 + h(z) \tag{B1}$$

Here $h(z)$ is assumed to be derivable but schematizes a highly varying pressure at the boundary of the two media: it decreases from $h(0) = P_1$ to $h(\psi) = 0$ and $h(z)$ is strongly variable in the interval $(0 - \psi)$. Consequently the space average $d\,h(x)/d\,z < 0$, while is nil around $z = 0$ and $z = \psi$. This initial condition can give some intuitive insight about a more realistic model for the two-rock border, where can happen complex phenomena as fine particles migrations, filter cake formation or local rock fractures (Merlani et al, 2011).

Replacing (B1) in equation (12) we have

$$\frac{\partial T}{\partial t} - k\frac{\partial^2 T}{\partial z^2} + \Delta\left(\frac{\partial T}{\partial z}\right)^2 + \Sigma \frac{d\,h}{d\,z}\frac{\partial T}{\partial z} + Y\left(\frac{d\,h}{d\,z}\right)^2 = 0 \tag{B2}$$

Mathematically this is a Burgers equation with a large positive inhomogeneous term $Y\left(\frac{dh}{dz}\right)^2$ and a drift term $\Sigma \frac{dh}{dz}\frac{\partial T}{\partial z}$. This problem must be treated numerically. A different possibility is that $Y\left(\frac{dh}{dz}\right)^2$ is a very large term, namely $h(z)$ is really highly variable function. Disregarding the other presumably smaller terms, an estimate at first order for space-averaged quantities therefore is

$$\frac{\partial T}{\partial t} + Y\left(\frac{dh}{dz}\right)^2 \approx 0 . \tag{B3}$$

This very simple case gives an elementary solution, approximately

$$T \approx T_0 - Y\left[\left(\frac{dh}{dz}\right)^2\right]t . \tag{B4}$$

**Appendix C. The structure of the fronts**

We here analyze some properties of the Burgers-like equation: this is not a formal mathematical demonstration but just an intuitive but exact sketch. Consider in general the equation

$$\frac{\partial T}{\partial t} = D\frac{\partial^2 T}{\partial z^2} + M\left(\frac{\partial T}{\partial z}\right)^2 + N\left(\frac{\partial T}{\partial z}\right), \tag{C1}$$

with $D$, $M$ and $N$ constants. We call $Q = \partial T/\partial z$ and by $z$-deriving (C1) we have

$$\frac{\partial Q(T)}{\partial t} - D\frac{\partial^2 Q(T)}{\partial z^2} - 2M\frac{\partial Q^2(T)}{\partial z} - \frac{\partial Q(T)}{\partial z} = 0 . \tag{C2}$$

Assuming that $T = T_0$ is constant in a small region around $z \approx a$ and $T = T_0 + T_I$ is again constant around $z \approx b$ we thus have that $Q(a) = Q(b) = 0$ in the above two small peripheral regions. In turn another $z$-derivative of (C2) gives that in small regions around $z = a$ and $z = b$ one has $\frac{\partial^2 Q}{\partial z^2} = M^2 \frac{\partial (Q^2)}{\partial z} = \frac{\partial Q}{\partial z} = 0$. Once integrated between $a$ and $b$ the relation (C 1) thus gives

$$\int_a^b \frac{\partial Q}{\partial t} dz = \frac{\partial}{\partial t}\int_a^b Q\, dz = \frac{\partial}{\partial t}[T(b) - T(a)] = 0 , \tag{C3}$$

that implies that $T(b) - T(a) = T_0 + T_I - T_0 = T_I = const$. This implies $T(b) = T(a) + T_I$.

If the solution of (C1) is growing like a polynomial $z, z^2,....$ in the $a - b$ interval and we fix that $a = 0$ and $b = z_B$, to satisfy the equation (C3) we must assume $T(z_B, t) = T_0 + T_I$, in particular in the limit $t \to 0$ and $z \to 0$.

**Appendix D. The effect of a filter cake formation.**

We now discuss another practical application of equations (24) - (29). We consider that during the transient crossing some small space scale, minerals and fine particle migrations can arrive within a short delay $d$ and affect the boundary layer permeability (Merlani et al., 2001). This can happen for very fine particles flowing into a solid matrix. Such flows would affect mainly the matrix porosity $K_f$ (and therefore $\alpha^*$, $k^*$ and $B$) since from (2) and (3) their effect on the porosity looks less important.

Obviously, during their migration, such fine particles deposit on the pore surfaces and therefore their volume $\gamma$ has to decrease. Calling $l$ the spatial rate of such decrease, one has

$$d\gamma/dz + l\gamma = 0, \text{ namely } \gamma(z) = \gamma_0 e^{-lz} \tag{D1}$$

Following Civan (1998), the effect on $K_f$ of such migration is assumed to be proportional to the volumetric flux of particles into the matrix. If a space averaging is an acceptable approximation for such migrations, their overall effect in the time interval $(d, \sqrt{4|\Delta T_1|\theta})$ is (Civan, 1998)

$$K_f \to K_f \left(1 + q \int_d^{\sqrt{4/\Delta/T_I\theta}} \gamma U \, dt \right) \approx K_f \left[1 + Q \ln\left(\frac{\sqrt{4/\Delta/T_I\theta}}{d}\right)\right] \tag{D2}$$

for a suitable parameter $q$ depending on the pore and particle dimensions. In (D2) the permeability is varied and the consequent novel velocity is

$U = -(K_f + \alpha) z / (2\mu \Delta \theta(t))$ and $Q = -q \int (K_f + \alpha)/(2\Delta\mu) \gamma_0 e^{-lz} z \, dz$.

**References:**

Bejan A., *Convection heat transfer*, John Wiley (1984).

Bonafede M., Hot fluid migration: an efficient source of ground deformation: application to the 1982-1985 crisis at Campi Flegrei-Italy, *J. Volc. Geotherm. Res.,* 48(1-2), 187-198 (1991).

Bonafede M., Mazzanti M., Hot fluid migration in compressible saturated porous media, *Geoph. Jour. Int.,* 128, 383-398 (1997).

Chester J.S., Chester, F.M., Kronenberg, A.K., Fracture surface energy of the Punchbowl fault, San Andreas system. *Nature,* 437.7055, 133-136 (2005).

Chirkov A.M., Radon as a possible criterion for predicting eruptions as observed at Karymsky volcano, *Bull. Volcanol.,* 39(1), 126-131 (1975).

Detournay E., Garagash, D. I., The near-tip region of a fluid-driven fracture propagating in a permeable elastic solid. *Journal of Fluid Mechanics*, 494, 1–32 (2003).

Fisher M.K., Davidson, B.M., Goodwin, A.K., Fielder, E.O., Buckler, W.S. and Steinberger, N.P., Integrating fracture mapping technologies to optimize stimulation in the Barnett shale, SPE Meeting, San Antonio, Texas, Expanded Abstract, SPE77411 (2002).

Garcia R., Natale, G., M. Monnin, and Seidel, J.L., Shock wave radon surface signals associated with the upsurge of *T-P* solitons in volcanic systems, *J. Volcan. Geotherm. Res.,* 96, 15-24 (2000).

Gangi A. F., Variation of whole and fractured porous rock permeability with confining pressure, *International Journal of Rock Mechanics and Mining Sciences,* 15(5), 249–257 (1978).

Garra R., Salusti, E., Droghei, R., Memory Effects on Nonlinear Temperature and Pressure Wave Propagation in the Boundary between Two Fluid-Saturated Porous Rocks, *Advances in Mathematical Physics*, Volume 2015, ID 532150, 6 (2015).

Gross D., Seeling T., *Fracture Mechanics*, Springer Verlag (2006).

Hummel N., Pressure dependent hydraulic transport as a model for fluid induced heartquakes, *Dissertation Freien Universitat Berlin* (2013).

McTigue D.F., Thermoelastic response of fluid-saturated rock, *J. Geoph. Research*., 91, 9533-9542 (1986).

Merlani A.L., Natale, G. and Salusti, E., Fracturing processes due to nonlinear shock waves propagating in fluid-saturated porous rocks, *Jour. Geoph. Research.* 106, 6 (2001).

Merlani A. L., Salusti, E. and Violini, G., Non-linear waves of fluid pressure and contaminant density in swelling shales, *Journal of Petroleum Science and Engineering* 79(1-2), 1-9 (2011).

Moore N.J., and Gunderson, R.P., Fluid inclusion and isotopic systematic of an evolving magmanic-hydrothermal system, *Geochim. Cosmochim. Acta*, 59, 3887-3907 (1995).

Natale G., The Effect of Fluid Thermal Expansivity on Thermo-mechanical Solitary Shock Waves in the Underground of Volcanic Domains, *Pure and applied geophysics*, 152(2), 193-211 (1998).

Natale G. and Salusti, E., Transient solutions for temperature and pressure waves in fluid-satured porous rocks, *Geophys. J. Int.*, 124, 649-656 (1996).

Natale G., E. Salusti, and A. Troisi, Rock deformation and fracturing processes due to nonlinear shock waves propagating in hyperthermal fluid-pressurized domains, *J. Geophys. Res.*, 103, 15,325-315,338 (1998).

Natale G., E. Salusti, and F. Tonani, Management of dry steam producing geothermal systems and volcanic explosion forecast: (*T, P*) solitary waves, a new potential source of significant information, paper presented at *24th Geothermal Reservoir Engineering, Stanford Univ., Stanford, Calif., Jan. 25-27* (1999).

Notsu K., H. Wakita, G. Igarashi, and T. Sato, Hydrological and geochemical changes related to the 1989 seismic and volcanic activities of the Izu Peninsula, *J. Phys. Earth*, 39, 245-254 (1991).

Philipp S., Afşar, F., Gudmundsson, A., Effects of mechanical layering on hydrofracture emplacement and fluid transport in reservoirs, *Frontiers in Earth Science* 1(4), 19 (2013).

Rempel A.W., Rice, J.R., Thermal pressurization and onset of melting in fault zones, *Journal of Geophysical Research: Solid Earth* 111.B9 (2006).

Rice J.R., Heating and weakening of faults during earthquake slip, *J. Geophysical Research*, 111 (B5), B05311, 29(2006).

Rice J.R., Cleary, M.P., Some basic stress-diffusion solutions for fluid saturated elastic media with compressible constituents, *Rev. Geophys.*, 14, 227–241 (1976).

Shapiro S.A. and Dinske, C., Fluid-induced seismicity: Pressure diffusion and hydraulic fracturing, *Geophysical Prospecting*, 57(2), 301–310 (2009).

Shapiro S. A., Dinske, C. and Rothert. E., Hydraulic-fracturing controlled dynamics of microseismic clouds. *Geophysical Research Letters,* 33.14 (2006).

Yilmaz O., Nolen-Hoeksema, R.C. and Nur,A. Pore pressure profiles in fractured and compliant rocks, *Geophysical prospecting,* 42.6: 693-714 (1994).

Whitham G.B., *Linear and Nonlinear Waves,* Wiley-Interscience, New York-London-Sydney, New York (1974).

Zencher F., Bonafede, M. and Stefansson, R., Near-lithostatic pore pressure at seismogenic depths: a thermoporoelastic model, *Geophys. J. Int.*, 166, 1318-1334 (2006).

**Figure Captions**

Fig 1. An intuitive sketch of the rock system.

Fig 2. Relation between strain σ and forcing ε (pressure and temperature) in 1-D problems

Fig 3. Solutions of the Burger equation for various **R.**

Fig. 4. Solutions for a positive *Δ*

Fig. 5. Solutions for a negative *Δ*

Fig. 6. Novel time $\theta$ for a pressure dependent permeability

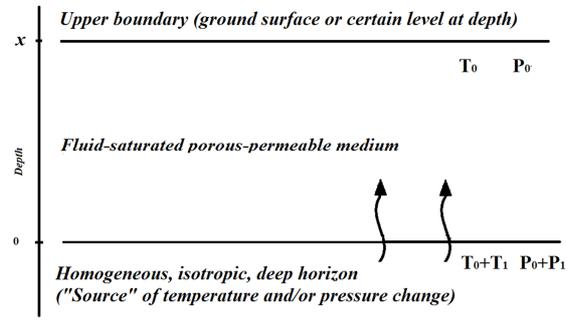

Fig 1. An intuitive sketch of the rock system.

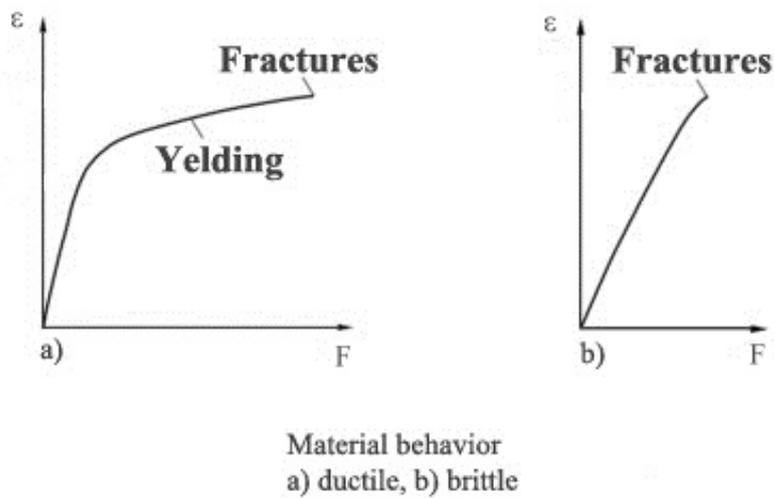

Fig 2. Relation between strain ε and forcing F (pressure and temperature) in 1-D problems

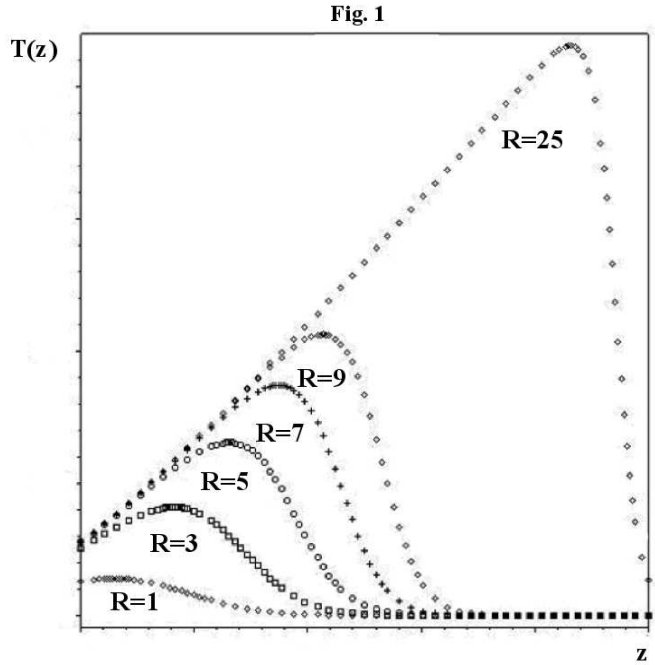

Fig 3. Solutions of the Burgers equation for various **R**.

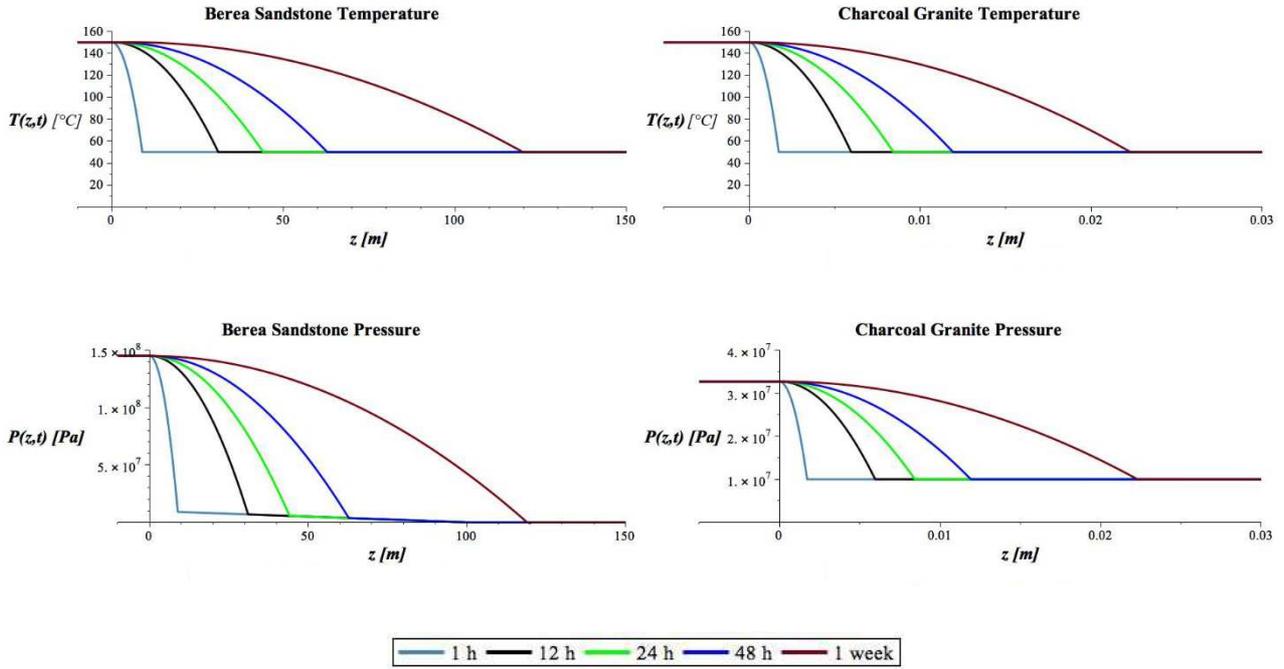

Fig. 4. Solutions for a positive $\varDelta$

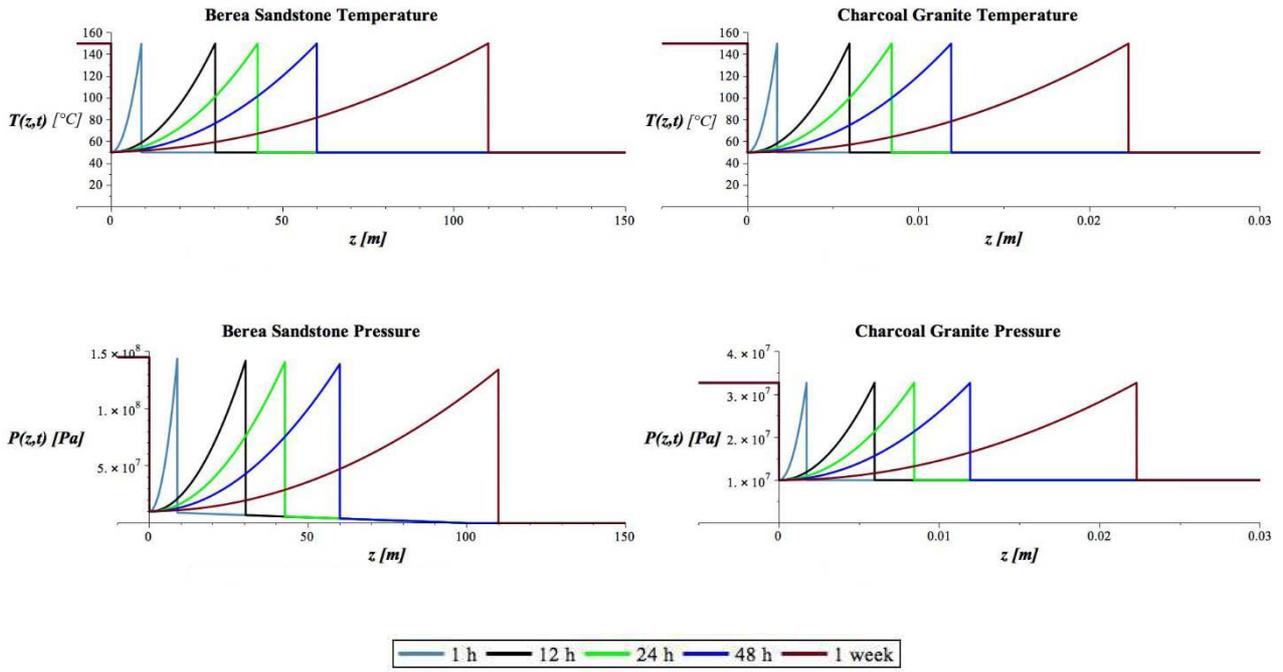

Fig. 5. Solutions for a negative $\Delta$

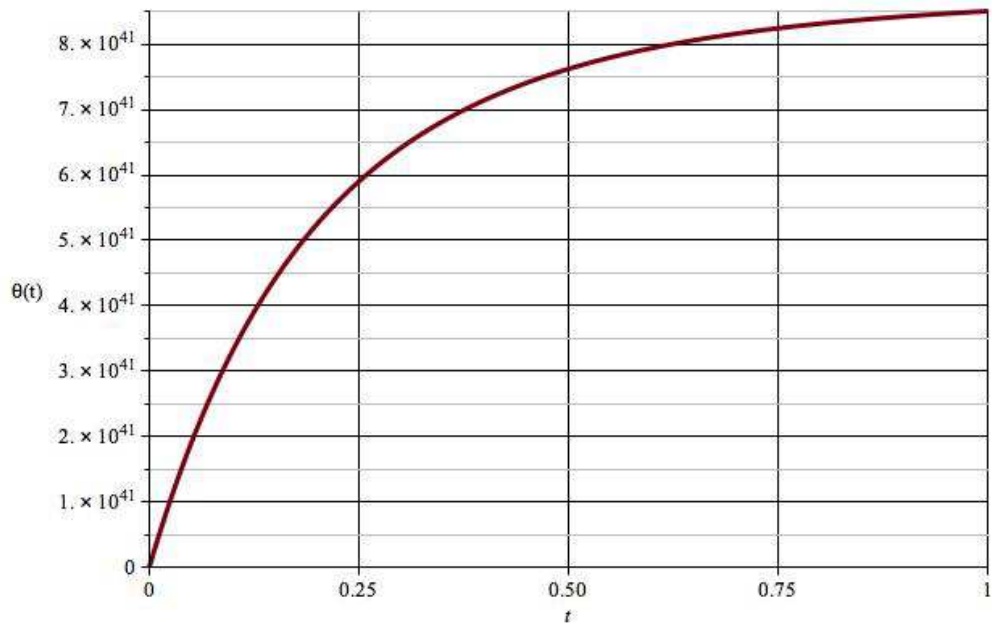

Fig. 6. Novel time $\theta$ for a pressure dependent permeability.

TABLES

TABLE 1

| Material property | Abyssal Red Clay | Berea Sandstone | Ruhr Sandstone | Weber Sandstone | Westerly granite | Charcoal granite | Tennessee Marble |
|---|---|---|---|---|---|---|---|
| $G$ | $7.0 \times 10^{4}$ | $6.0 \times 10^{9}$ | $1.2 \times 10^{10}$ | $1.2 \times 10^{10}$ | $1.6 \times 10^{10}$ | $1.8 \times 10^{10}$ | $2.4 \times 10^{10}$ |
| $B^*$ | $9.6 \times 10^{-1}$ | $6.2 \times 10^{-1}$ | $3.9 \times 10^{-1}$ | $4.0 \times 10^{-1}$ | $8.5 \times 10^{-1}$ | $2.3 \times 10^{-1}$ | $1.7 \times 10^{-1}$ |
| $v^*$ | $4.8 \times 10^{-1}$ | $2.0 \times 10^{-1}$ | $1.5 \times 10^{-1}$ | $1.5 \times 10^{-1}$ | $2.5 \times 10^{-1}$ | $2.7 \times 10^{-1}$ | $2.5 \times 10^{-1}$ |
| $v_u^*$ | $5.0 \times 10^{-1}$ | $3.3 \times 10^{-1}$ | $2.9 \times 10^{-1}$ | $2.2 \times 10^{-1}$ | $3.4 \times 10^{-1}$ | $3.0 \times 10^{-1}$ | $2.7 \times 10^{-1}$ |
| $\alpha_f$ | $3.0 \times 10^{-4}$ | $1.1 \times 10^{-3}$ | $9.8 \times 10^{-4}$ | $1.0 \times 10^{-3}$ | $1.0 \times 10^{-3}$ | $1.0 \times 10^{-3}$ | $1.0 \times 10^{-3}$ |
| $\alpha_m$ | $3.0 \times 10^{-5}$ | $3.0 \times 10^{-5}$ | $3.0 \times 10^{-5}$ | $3.0 \times 10^{-5}$ | $2.4 \times 10^{-5}$ | $2.4 \times 10^{-5}$ | $1.0 \times 10^{-5}$ |
| $K_f$ | $3 \times 10^{-16}$ | $2 \times 10^{-13}$ | $2 \times 10^{-16}$ | $1 \times 10^{-15}$ | $4 \times 10^{-19}$ | $10^{-19}$ | $10^{-19}$ |
| $K_T$ | 1.0 | 3.3 | 2.9 | 3.0 | 3.0 | 3.0 | 2.9 |
| $\varphi$ | $7 \times 10^{-1}$ | $2 \times 10^{-1}$ | $6 \times 10^{-2}$ | $6 \times 10^{-2}$ | $1 \times 10^{-2}$ | $2 \times 10^{-2}$ | $2 \times 10^{-2}$ |
| $\rho_m c_m$ | $4.0 \times 10^{6}$ | $3.0 \times 10^{6}$ | $3.0 \times 10^{6}$ | $2.7 \times 10^{6}$ | $2.7 \times 10^{6}$ | $2.6 \times 10^{6}$ | $1.0 \times 10^{9}$ |

Here $K_T$ is the average thermal conductivity, $K_f$ is the permeability, $\rho_m$ is the matrix density, $c_f$ is the fluid heat capacity and $c_m$ is that of the rock, $B^*$ is the Skempton parameter, $G$ is the shear modulus, $v^*$ the drained Poisson ratio and $v^*_u$ the undrained Poisson ratio, $\alpha_m$ ($\alpha_f$) the volumetric thermal expansion coefficient for the solid (fluid), $K_f$ is the medium permeability, $\phi$ the porosity, $c_f \rho_f \approx 7 \times 10^5$ and $\mu \approx 4 \times 10^{-3}$ in SI is the fluid viscosity.

*TABLE 2*

| Material property | Abyssal Red Clay | Berea Sandstone | Ruhr Sandstone | Weber Sandstone | Westerly granite | Charcoal granite | Tennessee Marble |
|---|---|---|---|---|---|---|---|
| $k*$ | $1.3 \times 10^{-7}$ | $4.2 \times 10^{-1}$ | $2.9 \times 10^{-4}$ | $2.5 \times 10^{-3}$ | $5.8 \times 10^{-6}$ | $3.1 \times 10^{-7}$ | $3.2 \times 10^{-7}$ |
| $\alpha*$ | $3.3 \times 10^{-7}$ | $5 \times 10^{4}$ | $62.5$ | $535$ | $1.7$ | $1 \times 10^{-1}$ | $5.7 \times 10^{-2}$ |
| $\alpha$ | $3 \times 10^{2}$ | $1.4 \times 10^{6}$ | $2.5 \times 10^{5}$ | $5 \times 10^{5}$ | $4.7 \times 10^{5}$ | $2.3 \times 10^{5}$ | $2.4 \times 10^{5}$ |
| $B$ | $3.1 \times 10^{-14}$ | $1.4 \times 10^{-11}$ | $1.2 \times 10^{-14}$ | $6.8 \times 10^{-14}$ | $2.6 \times 10^{-17}$ | $6.8 \times 10^{-18}$ | $1.8 \times 10^{-20}$ |
| $D$ | $5.9 \times 10^{-7}$ | $1.3 \times 10^{-6}$ | $1 \times 10^{-6}$ | $1.2 \times 10^{-6}$ | $1.1 \times 10^{-6}$ | $1.2 \times 10^{-6}$ | $3 \times 10^{-9}$ |
| $Y*$ | $4.4 \times 10^{-20}$ | $2 \times 10^{-17}$ | $1.8 \times 10^{-20}$ | $9.7 \times 10^{-20}$ | $3.7 \times 10^{-23}$ | $9.8 \times 10^{-24}$ | $2.6 \times 10^{-26}$ |
| $\|\Delta\|$ for $Y*$ positive | $9.4 \times 10^{-12}$ | $5.4 \times 10^{-5}$ | $4 \times 10^{-9}$ | $5.8 \times 10^{-8}$ | $2 \times 10^{-11}$ | $2 \times 10^{-12}$ | $5.8 \times 10^{-15}$ |
| $\|\Sigma\|$ for $Y*$ positive | $3.1 \times 10^{-14}$ | $6.7 \times 10^{-11}$ | $2.1 \times 10^{-14}$ | $1.6 \times 10^{-13}$ | $6.1 \times 10^{-17}$ | $1.1 \times 10^{-17}$ | $3 \times 10^{-20}$ |
| $\Delta/D = \mathbf{R}/T_1$ | $1.6 \times 10^{-5}$ | $4.2 \times 10^{1}$ | $4 \times 10^{-3}$ | $5 \times 10^{-2}$ | $1.9 \times 10^{-5}$ | $1.8 \times 10^{-6}$ | $2 \times 10^{-6}$ |
| $V = \Sigma \Gamma/\psi$ ($\Gamma = 10^{7}$; $\psi = 100$) | $3.1 \times 10^{-9}$ | $6.7 \times 10^{-6}$ | $2.1 \times 10^{-9}$ | $1.6 \times 10^{-8}$ | $6.1 \times 10^{-12}$ | $1.1 \times 10^{-12}$ | $3 \times 10^{-15}$ |
| $V = \Sigma \Gamma/\psi$ ($\Gamma = 10^{7}$; $\psi = 10^{-6}$) | $3.1 \times 10^{-1}$ | $6.7 \times 10^{2}$ | $2.1 \times 10^{-1}$ | $1.6$ | $6.1 \times 10^{-4}$ | $1.1 \times 10^{-4}$ | $3 \times 10^{-7}$ |

Characteristic parameters in SI for Abyssal Red Clay saturated with liquid water are estimated in (McTigue, 1986), the Berea sandstone and Rhur sandstone for supercritical water in Bonafede (1991). The values of the other rocks are from Merlani et al. (2001). Considering the difficulty of estimating these rock properties *in loco,* we give only the orders of magnitude of the above quantities, but the uncertainties can be large.